\documentstyle[emulateapj,psfig]{article}
\newcommand{\be}{\begin{equation}}
\newcommand{\ba}{\begin{eqnarray}}
\newcommand{\ee}{\end{equation}}
\newcommand{\ea}{\end{eqnarray}}  
\newcommand{\etal}{et al.\ }
\def\gtsima{$\; \buildrel > \over \sim \;$}
\def\ltsima{$\; \buildrel < \over \sim \;$}
\def\gsim{\lower.5ex\hbox{\gtsima}}
\def\lsim{\lower.5ex\hbox{\ltsima}}
\def\simgt{\lower.5ex\hbox{\gtsima}}
\def\simlt{\lower.5ex\hbox{\ltsima}}
\def\simpr{\lower.5ex\hbox{\prosima}}

\def\sngg{PPSNe }
\def\snggo{PPSNe}
\def\msun{\,{\rm M_\odot}}

\def\eg{{\frenchspacing\it e.g., }}
\def\E3{{\cal E}_{\rm g}^{III}}

\submitted{}

\begin{document}
\title{The Detectability of Pair-Production Supernovae at ${
z\lsim 6}$} \author{Evan Scannapieco\altaffilmark{1}, Piero
Madau\altaffilmark{2},  Stan Woosley\altaffilmark{2},  
\qquad \qquad \qquad 
\qquad \qquad \qquad \qquad \qquad \qquad 
Alexander Heger\altaffilmark{2,3,4}, \&  Andrea Ferrara\altaffilmark{5}}
\altaffiltext{1}{ Kavli Institute for Theoretical Physics, Kohn Hall,
UC Santa Barbara, Santa Barbara, CA 93106.}
\altaffiltext{2}{Department of Astronomy \& Astrophysics, UC Santa
Cruz, Santa Cruz, CA, 95064.}  \altaffiltext{3}{Theoretical
Astrophysics Group, T6, MS B227, Los Alamos National Laboratory, Los
Alamos, NM, 87545.}  \altaffiltext{4}{Enrico Fermi Institute, The
University of Chicago,  5640 S.\ Ellis Ave., Chicago, IL 60637.}
\altaffiltext{5}{SISSA/International School for Advanced Studies, Via
Beirut 4, I-34914, Trieste, Italy.}

\begin{abstract}

Nonrotating, zero metallicity stars with initial masses $140\lsim
M_\star \lsim 260\,\msun$ are expected to end their lives as
pair-production supernovae (\snggo), in which an electron-positron
pair-production instability triggers  explosive nuclear burning.
Interest in such stars has been rekindled by recent theoretical
studies that suggest primordial molecular clouds preferentially form
stars with these masses.  Since metal enrichment is a local process,
the resulting \sngg could occur over a broad range of redshifts, in
pockets of metal-free gas.  Using the implicit hydrodynamics code
KEPLER, we have calculated a set of PPSN light curves that addresses
the theoretical uncertainties and allows us to assess observational
strategies for finding these objects at intermediate redshifts.  The
peak luminosities of typical \sngg are only slightly greater than
those of Type Ia, but they remain bright much longer ($\sim$ 1 year)
and have hydrogen lines.  Ongoing supernova
searches may soon be able to limit the contribution of these very
massive stars to $\lsim 1\,$\% of the total star formation rate
density out to $z \approx 2,$ which already provides useful
constraints for theoretical models. The planned {\em Joint Dark Energy
Mission}  satellite  will be able to extend these limits out to $z
\approx 6$.

\end{abstract}


\section{Introduction}

Determining the cosmological evolution of the elements and elucidating
the processes responsible for their creation are key aspects of
understanding the history of the universe.  A wide spread in
abundances is observed in stars, but the lowest metallicity galaxies
studied so far are enriched to substantial values of $\sim 0.02
Z_\odot$ (\eg Searle \& Sargent 1972), and the diffuse medium in
galaxy clusters is enriched to levels above 1/3 $Z_\odot$ (\eg Renzini
1997).  Similarly, quasar absorption line studies have uncovered heavy
elements in even the lowest-density regions of the intergalactic
medium (IGM) (\eg Schaye \etal 2003; Aracil \etal 2004) at levels that
are roughly constant out to the highest redshifts probed (Songaila
2001; Pettini \etal 2003).  And though direct observations of
metal-poor halos stars have uncovered a number of unusual abundance
patterns (\eg Ryan, Norris, \& Beers 1996; Christlieb \etal 2002;
Cayrel \etal 2004), no metal-free star has been detected to date.

Thus a clear signature of the transition from a metal-free to an
enriched universe has so far remained unobserved. Furthermore, there
are good reasons to believe that this transition may have marked a
drastic change in the characteristics of star formation.  Recent
theoretical studies have shown that collapsing primordial clouds
fragmented into clumps with typical masses $\sim 10^3 \msun$ (\eg
Nakamura \& Umemura 1998; Abel, Bryan, \& Norman 2000; Bromm \etal
2001) and that accretion onto proto-stellar cores within these clouds
was very efficient (Ripamonti \etal 2002; Tan \& McKee 2004).  The
implication is that the initial mass function (IMF) of metal-free
``Population III'' stars may have been biased to higher masses than
observed today.

Since only a trace level of heavy elements will drastically increase
fragmentation, the formation of very massive stars  (VMS) is unlikely
beyond a ``critical metallicity'' of $Z_{\rm crit} \sim 10^{-4}$ (\eg
Schneider \etal 2002; Bromm \& Loeb 2003).  This means that the metals
produced by the first few supernovae (SNe) were sufficient to halt
very massive star formation, implying that this mode may have only
operated at extremely high-redshifts $z \gsim 15$ (\eg Mackey, Bromm,
\& Hernquist 2003).  Yet, such simple estimates ignore the substantial
propagation times necessary for heavy elements to be mixed into
pristine material.  In fact, more detailed analyses show that it is
almost impossible to raise the IGM metallicity above $Z_{\rm crit}$ in
a homogeneous way (Scannapieco, Schneider, \& Ferrara 2003, hereafter
SSF03).  Thus primordial metal enrichment was essentially a local
process, which began in the densest ``high-sigma'' regions and
gradually spread into to the more common lower-density regions as time
progressed.  This inhomogeneous distribution parallels models of
further IGM enrichment by second-generation objects (\eg Scannapieco,
Ferrara, \& Madau 2002; Theuns \etal 2002; Fujita \etal 2004; Cen,
Nagamine \& Ostriker 2004) and raises the possibility that metal-free
star formation may extend well into the observable range, in pockets
of gas where metals have not yet propagated.

Such primordial star-forming regions would naturally be isolated.
Similarly, metal-free objects would tend to be just large enough for
gravitational collapse to overcome the IGM thermal pressure, but small
enough to have no direct progenitors above a cosmological Jeans mass.
After reionization, such pristine star
formation would be fundamentally different than that taking place in
minihalos at very high redshifts, which depends sensitively on the
presence of initial $H_2$ (\eg O'Shea \etal 2005).  In small objects,
molecular hydrogen is easily photodissociated by  11.2-13.6 eV
photons, to which the universe is otherwise transparent.   Thus the
emission from the first stars quickly destroyed all avenues for
cooling by molecular line emission (Dekel \& Rees 1987; Haiman, Rees,
\& Loeb 1997; Ciardi, Ferrara, \& Abel 2000).  This quickly raised the
minimum virial temperature necessary to cool effectively to
approximately $10^4$ K, although the precise value of this transition
is the subject of debate  (\eg Glover \& Brand 2001 Yoshida \etal
2003) and is somewhat dependent on the level of the high-redshift
X-ray background (Haiman, Rees, \& Loeb  1996; Oh 2001; Machack,
Bryan, \& Abel 2003).

Later metal-free star formation is much more straightforward, and is
confined to halos with $T_{\rm vir} \geq 10^4$ K.  In  this case
efficient atomic line cooling establishes a dense locally-stable disk,
within which non-equilibrium free electrons catalyze $H_2$. Unlike in
less massive halos, $H_2$ formation in these objects is largely
impervious to feedback from external UV fields, due to the  high
densities achieved by atomic cooling (Oh \& Haiman 2002).
As discussed in SSF03, these lonely regions of late primordial star formation 
may have already been detected as a subgroup of high-redshift Lyman-alpha
emitters, although distinguishing these objects from Population II
stars is extremely difficult (Dawson \etal 2004).  

In this investigation we explore an alternative approach to search for
such VMS: the direct detection of the resulting SNe.  While many of
the $M_\star \geq 100 \msun$ stars would have died forming black
holes, progenitors within the $140-260 \msun$ mass range end their
lives in tremendously powerful pair-production supernovae (\snggo).
In these SNe, $e^+ e^-$ pair creation softens the equation state at
the end of central carbon burning, leading to rapid collapse, followed
by explosive burning up to the iron group (\eg Ober, El Eid, \& Fricke
1983; Bond, Arnett, Carr 1984; Heger \& Woosley 2002, hereafter HW02).
The more massive the star, the higher the temperature at bounce and
the heavier the elements that are produced by nuclear fusion.  In all
cases, the
star is completely disrupted and both the ejection energies ($\sim
3-100 \times 10^{51}$ ergs) and masses ($140-260 \msun$) are enormous,
leading to observed properties qualitatively different from those of
Type-Ia and typical core collapse SNe.  In turn, these properties,
such as bright UV colors, allow for efficient intermediate-redshift
searches for very massive star formation.

Recently, Wise \& Abel (2005) studied the feasibility of detecting
\sngg at $z \sim 20$ with the {\it James Webb Space Telescope}, using
extremely approximate lightcurves with luminosities that were taken to
be proportional to the total $^{56}$Ni mass.  A similar study was
carried out by Weinmann \& Lilly (2005), who also  estimated number
counts at $z=5$ using a single light curve that was taken from Heger
\etal (2002), which unfortunately contained a normalization
error.\footnote{The curves in this conference proceeding are
overluminous by a factor of $1+z.$}   Limited studies of lightcurves
from \sngg have also been carried out by Woosley \& Weaver (1982), who
considered progenitors with only two masses and found features similar
to those described below, and by  Herzig \etal (1990) who modeled the
disruption of a 61 $\msun$ Wolf-Rayet star, generating a relatively
faint  SN.  Here we use the KEPLER code to directly construct a suite
of PPSN models  that addresses the range of theoretical possibilities
and accounts for the detailed physical processes involved up until the
point at which the SN becomes optically thin.  These curves allow us
to identify the properties that most distinguish \sngg from Type Ia
and core collapse SNe, and discuss their detectability at $z \lesssim
6$ in current and future surveys.

The outline of this paper is as follows.  In \S 2 we use the implicit
hydrodynamical code KEPLER to construct a suite of detailed PPSN
progenitor  models. In \S 3 we generate simulated spectra and
lightcurves from these models, highlighting the key  features that
distinguish \sngg from other SNe.  In \S 4 we apply these lightcurves
to determine the constraints on \sngg    obtainable by existing and
planned SNe searches.  In \S 5 we discuss the  properties of the most
likely environments of \snggo, and we summarize our results in \S 6.

\begin{table*}
\begin{center}
\vspace{0.5mm}
\caption{Properties of PPSN progenitor models}
\vspace{3mm}
\begin{tabular}{l|cccccc} 
\hline
Model &   $M_{\rm He}$ ($\msun$)  &  $M_{\rm N}$ ($\msun$)
  & $M_{^{56}{\rm Ni}}$ ($\msun$) &  $R$ ($10^{13}{\rm cm}$)
  &  ${\cal E}_{\rm kin}$ ($10^{51}{\rm ergs}$) \\ \hline\hline

150-W   &  70   &  3.5(-4)  &  4.2(-2)  & 3.9    &  6.9   \\ 
150-I   &  46   &  1.1(-4)  &  6.3(-2)  & 16     &  9.2   \\ 
150-S   &  49   &  0.86     &  8.6(-2)  & 26     &  8.5  \\ 
200-W    &  97   &  2.7(-6)  &   3.3    &  0.68  &  29.5  \\ 
200-I    &  58   &  8.0(-6)  &   5.1    &  2.8   &  36.5  \\ 
200-S    &  89   &  0.34     &   2.2    &  29    &  29.1  \\ 
200-S2   &  78   &  4.75     &   0.82   &  20    &  18.7  \\ 
250-W    &  123  &  3.1(-6)  &   6.2    &  0.58  &  47.2  \\ 
250-I    &  126  &  9.1(-6)  &   32     &  4.0   &  76.7  \\ 
250-S    &  113  &  1.34     &   24.5   &  26    &  64.6  \\ 
\hline
\end{tabular}
\end{center}
\end{table*}

\section{Models of Pair-Production Supernovae}

The progenitors of \sngg are confined to a range of initial masses
that are large enough to collapse via the pair-production instability,
but small enough for this collapse to be reversed by nuclear burning.
In the non-rotating case, these limits are clearly defined and
well understood  numerically (Ober, El Eid, \& Fricke 1983; HW02; Heger
\etal 2003).  In stars above $\sim100 \msun$, after central helium
burning, the core reaches a sufficiently high temperature and entropy
to  produce $e^+e^-$ pairs, converting some of the internal energy of
the gas into rest mass, which softens the equation of state to an
adiabatic index below the critical value of 4/3 (Barkat, Rakavy, \&
Sack 1967; Bond, Arnett, \& Carr 1984).  The result is a runaway
collapse that leads to explosive burning in the carbon-oxygen
core. For stars with initial masses more than about 140 $\msun,$ the
energy released is sufficient to completely disrupt the star.

A shock moves outward from the edge of the core, initiating the
supernova outburst when it reaches the stellar surface.  Just above
the 140 $\msun$ limit, weak silicon burning occurs and only trace
amounts of radioactive $^{56}$Ni are produced.  It is this $^{56}$Ni
that powers the late-time supernova light curves.  The amount of
$^{56}$Ni produced increases in larger progenitors, and in 260 $\msun$
progenitors up to 50
$\msun$ may be synthesized, $\sim$ 100 times more than in a typical
Type Ia supernova.  For stars with masses above 260 $\msun,$ however,
the onset of photodisintegration in the center imposes an additional
instability that collapses most of the star into a black hole (Bond,
Arnett \& Carr 1984; HW02; Fryer, Woosley, \& Heger 2001).

Pair-production supernovae are among most powerful thermonuclear
explosions in the universe, with a total energies ranging from $3 \times
10^{51}$ ergs for a 140 $\msun$ star (64 helium $\msun$ core) to
almost $100 \times 10^{51}$ ergs for a 260 $\msun$ star (133 $\msun$
helium core; HW02).  We use the implicit hydrodynamical code KEPLER
(Weaver, Zimmerman, \& Woosley 1978) to model the entire evolution of
the star and the resulting light curves.  Since KEPLER only implements
gray diffusive radiation transport with approximate deposition of
energy by gamma rays from radioactive decay of $^{56}$Ni and $^{56}$Co
(Eastman, Woosley, \& Weaver 1993), the light curves obtained can only
be followed as long as the SN is optically thick an there is a
reasonably well-defined photosphere.  Also during the very earliest
stages of the supernova (shock break out), KEPLER underestimates the
color temperature by about a factor 2 (Blinnikov \etal 2003).
Finally, we employ a mixing procedure to simulate, in 1-D, the effects
of mixing by Rayleigh-Taylor instabilities after the supernova shock,
which was calibrated using spectra of observed SN types.  The
main effect of this mixing for the light curves is the spreading of
some $^{56}$Ni into the envelope.

A supernova can be bright either because it makes a lot of radioactive
$^{56}$Ni (as in Type Ia supernovae) or because it has a large low
density envelope and large radius (as in bright Type II supernovae).
More radioactivity gives more energy at late times when the supernova
has expanded enough to become transparent (R $\sim$ 10$^{15}$ cm).  A
larger initial radius means that the star experiences less adiabatic
cooling as it expands to that radius, resulting in  a higher
luminosity at early times.  Here, the most important factors in
determining the resulting light curves are the mass of the progenitor
star and the efficiency of dredge-up of carbon from the core into the
hydrogen envelope during or at the end of central helium burning.  The
specifics of the physical process encountered here are unique to
primordial stars.  Lacking initial metals, they have to produce the
material for the CNO cycle themselves, through the synthesis of
$^{12}$C by the triple-alpha process.  Just enough $^{12}$C is
produced to initiate the CNO cycle and bring it into equilibrium: a
mass fraction of $10^{-9}$ when central hydrogen burning starts, and a
mass fraction $\sim 10^{-7}$ during hydrogen-shell burning.

At these low values the entropy in the hydrogen shell remains barely
above that of the core, and the steep entropy gradient at the upper
edge of the helium core that is typical for metal-enriched
helium-burning stars is absent. The high masses of the stars
also imply that a large fraction of the pressure is due to
radiation. This is well known to facilitate convection.  For these
reasons, the central convection zone during helium burning can get
close, nip at, or even penetrate the hydrogen-rich layers.  Once
such mixing of high-temperature hydrogen and carbon occurs, the two
components burn violently, and even without this rapid reaction,
the hydrogen burning in the CNO cycle increases proportionately
to the additional carbon.  Thus mixing of material
from the helium-burning core, which has a carbon abundance of order
unity, is able to raise the energy generation rate in the
hydrogen-burning shell by orders of magnitude over its intrinsic
value.

\begin{figure*}[t]
\vspace{130mm}
\includegraphics{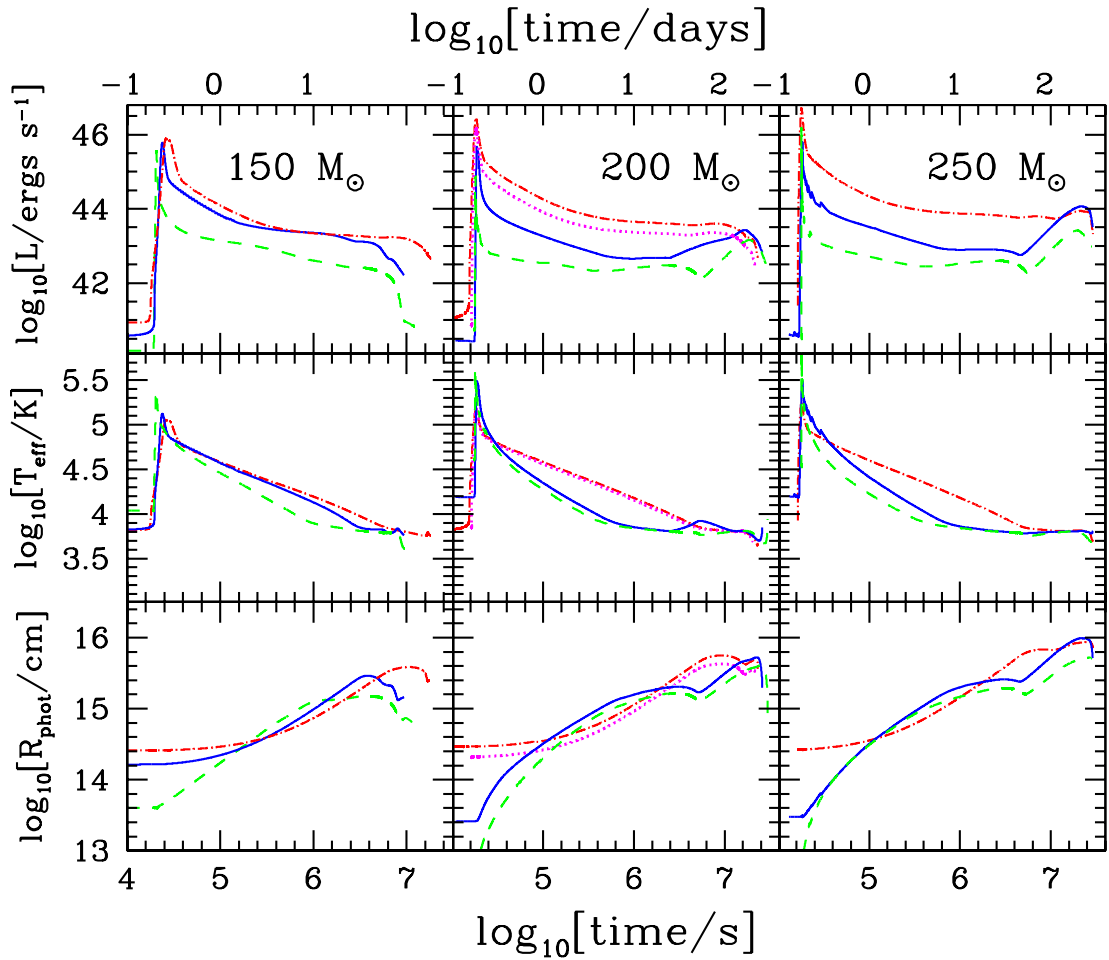}
\caption{\footnotesize 
Luminosities (top row), effective temperatures (center row),
and photospheric radii (bottom row) of \sngg for ten different
representative models.   Each sets of panels are labeled by the total
mass of the progenitor star, and models with weak (dashed),
intermediate (solid) and  strong (dot-dashed) convective overshoot
are shown.  Finally, the dotted lines in the central panels correspond
to model 200-S2.  See text and Table 1 for details.}
\label{fig:sne}
\end{figure*}

This mixing has two major effects on the PPSN progenitor: first, it
increases the opacity and energy generation in the envelope, leading
to a red-giant structure for the presupernova star, in which the radius
increases by over an order of magnitude.  Second, it decreases the
mass of the He core, consequently leading to a smaller mass of
$^{56}$Ni being synthesized and a smaller explosion energy.  The
former effect increases the luminosity of the supernova, especially at
early times, while the latter effect can weaken it.  Just how much
mixing occurs is uncertain from stellar modeling at the present,
though it has been studied by several authors (\eg Heger, Woosley \&
Waters 2000; Marigo \etal 2001).
In the present paper we account for these uncertainties by employing
different values of convective overshooting -- whose presence is
clear, but whose quantitative interaction with the 
burning is unknown.

A suite of representative models is chosen to address the expected
range of presupernova models - from blue supergiant progenitors with
little or no mixing to well-mixed red hypergiants, which can have
pre-SN radii of 20 AU or more.   These models are summarized in Table
1, in which the names refer to the mass of the progenitor star
(in units of $\msun$) and the  weak (W), intermediate (I), or strong
(S) level of of convective  overshoot.  Note, that in all cases the
progenitor star is nonrotating, and all models employ the same
procedure to mix $^{56}$Ni into the envelope after the supernova shock.

The KEPLER code can be used to compute approximate light curves and
has been validated both against much more complex and realistic codes
such as EDDINGTON and observations of a prototypical Type II-P
supernova, SN 1969L (Weaver and Woosley 1980; Eastman et al 1994). Its
deficiency is that it is a single temperature code using flux-limited
radiative diffusion. It functions very well, however, in situations
where the light curve is recombination dominated (as on the plateau of
Type II-P supernovae), in the radioactive peaks of Type I supernova,
and in the tails of Type II-P supernovae.  Gamma-deposition has been
calibrated against a Monte Carlo code (Pinto and Woosley 1988) and the
code was successfully used to predict the behavior of SN 1987A
(Woosley, Pinto, \& Ensman 1988), though mixing was later added as an 
essential ingredient.  Because thermal equilibrium and black body spectra are
assumed, the bolometric luminosity is known much better than the color
photometry. Thus B and V magnitudes should be approximately correct, but U
magnitudes are much more uncertain.

The resulting luminosities, effective temperatures,  photospheric
radii are shown in Figure \ref{fig:sne}.  As the shock moves toward
the low-density stellar surface, its energy is deposited into
progressively smaller amounts of matter.  This results in high
velocities and temperatures when the shock reaches the stellar
surface, causing a pulse of ultraviolet radiation with a
characteristic timescale of a few minutes.  This ``breakout'' phase is
by far the most luminous and bluest phase of the PISN burst, but its
very short duration makes it difficult to use in observational
searches.  In fact, the analog of this phase in conventional SN has so
far only been indirectly detected in SN 1987A (Nayozhin
1994; Hamuy \etal 1988; Catchpole \etal 1988).

\begin{figure*}[t]
\vspace{135mm}
\includegraphics{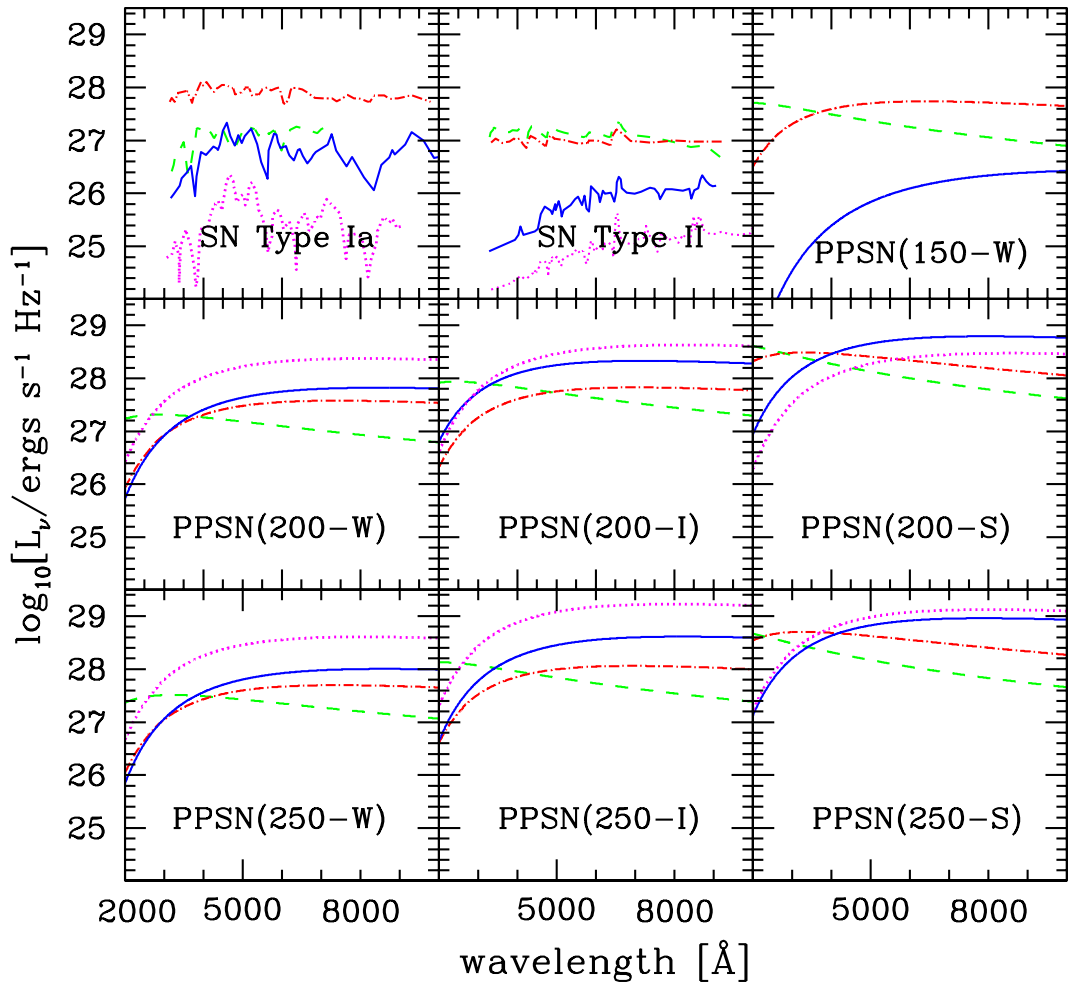}
\caption{\footnotesize 
Comparison of \sngg spectra with those of Type Ia and Type
II-P supernovae at 1 day (dashed), 10 day (dot-dashed), 100 days
(solid), and  200 days (dotted) since explosion.  The Type Ia spectra
were taken from observations  of SN 1994D, the Type II-P spectra were
taken  from observations of SN 1999em, and the \sngg spectra were
computed from black body models  with $T_{\rm color} = T_{\rm eff}.$
While the overall shape of our simplified \sngg spectra are similar to
those of SNe Type Ia and Type II, their late-time evolution is
markedly different, as seem most clearly in the 200  day curves.}
\label{fig:spectra}
\end{figure*}

Following breakout, the star expands with $R_{\rm phot}$ initially
proportional to time.  Though a small fraction of the outer mass may
move much faster, the characteristic velocity of the photosphere
during this phase is a modest $v = (2 KE/M)^{1/2} \sim (10^{53} {\rm
ergs}/200 \msun)^{1/2} \sim 5000$ km/s, because of the very large mass
participating in the explosion.  During the expansion, the
radiation-dominated ejecta  cool adiabatically, with $T$ approximately
proportional to $R^{-1},$ with an additional energy input from the
decay of $^{56}$Ni (if a significant mass was synthesized during the
explosion) and hydrogen recombinations (when $T \sim 10^4$K).  As the
scale radius for this cooling is the radius of the progenitor, the
temperatures and luminosities are substantially larger throughout this
phase in the cases with the strongest mixing.

After $\sim 50$ days, the energy input from $^{56}$Co decay
becomes larger than the remaining thermal energy (the initial thermal
energy deposited by the shock is mostly eaten away by the adiabatic
expansion).  At this time the energy deposited by $^{56}$Co in deeper
layers that were enriched in $^{56}$Ni can diffuse out and these
layers also become gradually more exposed as the outer parts of the
supernova ejecta recombine and become optically thin. For stars that
were compact to begin with, this can cause a delayed rise to the peak
of the light curve.  For stars with larger radii, the radioactivity
just makes a bright tail following the long plateau in emission from
the expanding envelope.  Eventually, even the slow-moving inner layers
recombine and there is no longer a well-defined photosphere.  At this
time the assumption of local thermodynamic equilibrium (LTE) breaks
down, and more detailed radiative transfer calculations are required,
which are beyond the scope of our modeling here. The SN is fainter and
redder during this phase, however, and thus difficult to detect at
cosmological distances in optical and near infrared (NIR) surveys.

\begin{figure*}[t]
\vspace{155mm}
\includegraphics{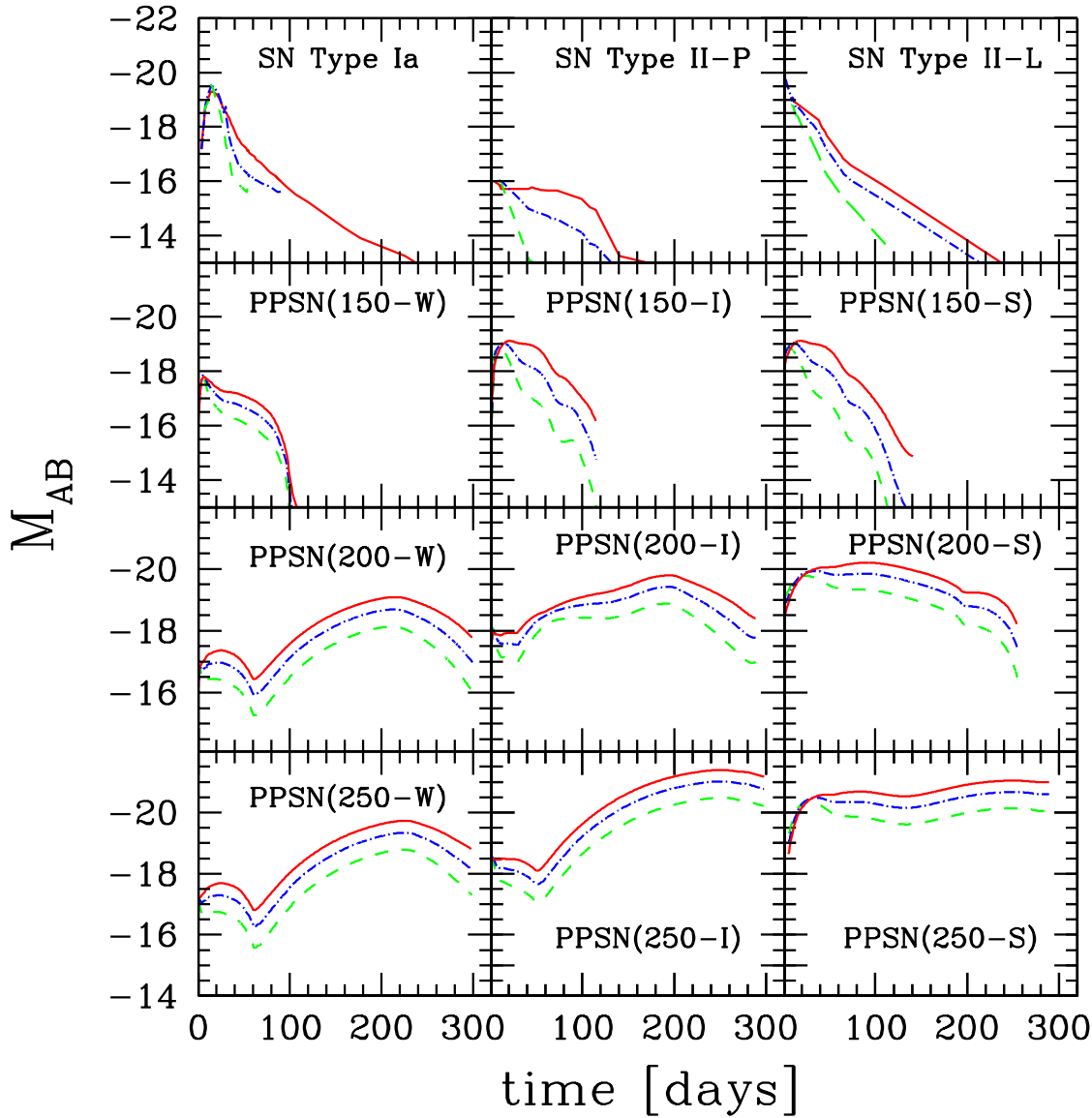}
\caption{\footnotesize 
Comparison of light curves of a SN Type Ia, a SN Type II-P,
a bright SN Type II-L, 
and \sngg models with varying progenitor masses.  In all cases 
the solid lines are absolute V-band AB magnitudes,
the dot-dashed lines are the absolute B-band AB magnitudes, and 
the dashed lines are the absolute U-band AB magnitudes.}
\label{fig:lightcurves}
\end{figure*}

\section{Spectral Properties and Lightcurves}

From the luminosities and effective temperatures in Figure
\ref{fig:sne}, we calculated approximate \sngg spectra, assuming a
black body distribution with the color temperature equal to the
effective temperature, that is the specific luminosity \be L_\nu(t) =
L(t) \frac{15 x^3}{\nu_0 \pi^4 (e^x-1)},  \ee  where $x \equiv
\nu/\nu_0$ and $\nu_0 = 2.1 \times 10^{14} \, T_{\rm 4,eff}$ Hz, with
$T_{\rm 4,eff} \equiv T_{\rm eff}/10^4$ K.  Recall that the peak
frequency in this case occurs at $5.9 \times 10^{14} \, T_{\rm 4,eff}$
Hz, which corresponds to a wavelength of $5100 \, T_{\rm 4,eff}^{-1}$
\AA.  The resulting spectra are plotted at representative times of 1,
10, 100, and 200 days in Figure \ref{fig:spectra}, in which we also
include comparison spectra from observations of Type Ia and Type II
(core collapse) SNe.  In particular, the Type Ia curves were taken
from observations of SN 1994D by Patat \etal (1996) and  Filippenko
(1997) as compiled by Filippenko (1999), at 1, 10 and 128 days, and
233 days since  explosion (assumed to be 13 days before B-max)
normalized to Patat \etal (1996) and Cappelaro \etal (1997) V-band
light curves at 1, 10,  100, and 200 days, with an assumed distance of
13.7 Mpc (see also Richmond \etal 1995).  The Type II curves, on the
other hand, were taken from spectra of the Type II-P SN 1999em by
Elmhamdi \etal (2003) at 9, 10, 113, and 168 days since explosion, again
normalized to their V-band light curves at 1, 10, 100, and 200 days, with
an assumed distance of 7.8 Mpc.

The most striking feature from this comparison is that despite
enormous kinetic energies of  $\sim 50 \times 10^{51}$ ergs, the peak
optical luminosities of \sngg are similar to those of other SNe, even
falling below the Ia and II curves in many cases.  This is because the
higher ejecta mass produces a large optical depth and most of the
internal energy of the gas is converted into kinetic energy by
adiabatic expansion.  Furthermore, the overall spectral shapes of the
PPSN curves are not unlike those of more usual cases.  In fact,
pair-production supernovae spend most of their lives in the same
temperature range as other SNe and therefore exhibit a similar range
in colors throughout their evolution.   Lastly, many of line features
in the observed spectra would also be present in  more detailed models
of \snggo.  In particular, because of their hydrogen envelopes, PPSN
spectra should contain hydrogen lines similar to those in Type-II SNe.
Clearly, then, \sngg will not be obviously distinguishable from their
more usual counterparts ``at first glance.''

A closer comparison between spectra, however, uncovers two key
features that are uniquely characteristic to \snggo.  The first of
these is a dramatically extended intrinsic decay time, which is
especially noticeable in the models with the 
strongest enrichment of CNO in the envelope. This
is due to the long adiabatic cooling times of supergiant progenitors,
whose radii are $\sim 20$ AU, but whose expansion velocities are
similar or even less than those of other SNe.  Second, \sngg are the
only objects that show an extremely late rise at times $\geq 100$
days.  This is due to energy released by the decay of $^{56}$Co, which
unlike in the Type Ia case, takes months to dominate over the internal
energy imparted by the initial shock.  In this case the feature is
strongest in models with the least mixing and envelope enrichment
during helium burning, as these have the largest helium cores and
consequently the largest $^{56}$Ni masses.  Note, however, that
neither of these features is generically present in all \snggo, and
both can be absent in smaller VMS that fail expand to large sizes
through dredge-up and do not synthesize appreciable amounts of
$^{56}$Ni.  In the 150-W case, for example, the luminosity decays
monotonically on a relatively short time scale, producing spectra not
dissimilar to the comparison Type II curves from SN 1999em.  In fact
this 150 $\msun$ SN shares many similarities with its smaller-mass
cousin: both are SNe from progenitors with radii $\sim 10^{13}$ cm and
in both $^{56}$Ni plays a negligible role.

In Figure \ref{fig:lightcurves} we examine the temporal evolution of
\sngg in more detail by plotting absolute AB light curves of our
models at three representative wavelengths: 5500 \AA, corresponding
to the central wavelength of the V-band; 4400 \AA, corresponding to
the B-band; and 3650 \AA, corresponding to the U-band.
We focus on blue wavelengths as it is features in these
bands that will be redshifted into the optical and NIR at cosmological
distances.  Again, for comparison, we also include in this figure
observed light curves for SN Type Ia and Type II.  
In the Type Ia case the curves are again taken from
observations of 1994D by Patat \etal (1996) and are supplemented at
late times by data from Cappelaro \etal (1997).   In the Type II case,
we consider both Type II-P and Type II-L  SNe.  Our Type II-P curves are
taken from observations of 1999em by Elmhamdi \etal (2003),
and our Type II-L curves are taken from observations of the
very bright supernova 1979C, as compiled in de Vaucouleurs \etal
(1981) and Barbon \etal (1982), with an assumed distance of 17.2 Mpc
(Freedman \etal 1994).  

From this point of view, the characteristic features of \sngg are even
more striking, and the relation with progenitor structure 
is clear.  Compact models with weak dredge-up are
dominated by a late-time $^{56}$Ni bump; extended hypergiant
progenitor models with strong mixing show a significant blue early
phase during the expansion followed by a longer, redder phase in which
the luminosity is roughly constant for $\sim 200$ days; and
intermediate models are a hybrid of the two.  Finally, in cases  such
as the compact 150-W blue supergiant model, in which dredge-up does
not occur and little $^{56}$Ni is synthesized, both of these phases
are absent, leaving an object remarkably reminiscent of more typical
core collapse SNe, as exemplified by the Type II-P lightcurves.
Similarly, models such as 150-I and 150-S which also synthesize very
little $^{56}$Ni, but are preceded by larger progenitors, are easily
confused with bright Type II-L SNe, particularly at  early times.   In
no case, however, do \sngg look anything like SNe Type Ia.  In
particular none of the  pair-production models display the long
exponential decay seen in the Type Ia curves, and all \sngg contain
hydrogen lines, arising from their substantial envelopes.

\begin{figure*}[t]
\vspace{95mm}
\includegraphics{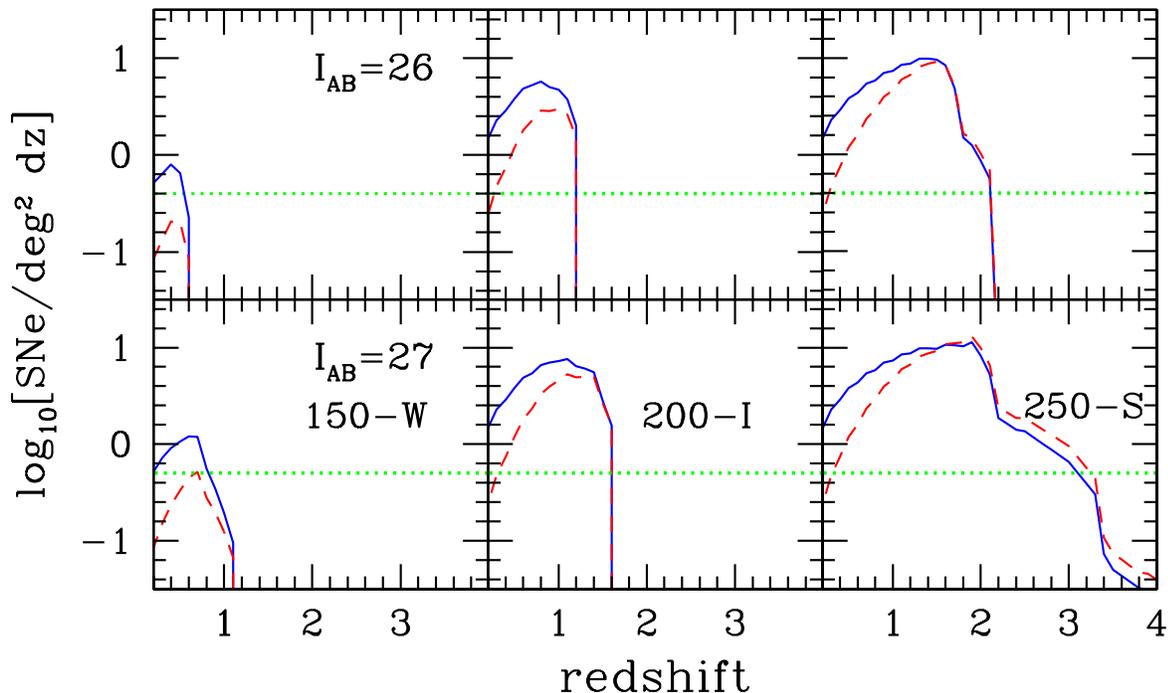}
\caption{\footnotesize 
Number of \sngg  per square degree per unit redshift above a
given I-band magnitude, assuming 0.001 $\msun$ yr$^{-1}$ Mpc$^{-3}$
(solid lines) or 1\% of the observed star formation rate density (dashed
lines).   The $I_{AB} = 26$ cut taken in  the upper rows approximately
corresponds to the magnitude limit of the  Institute for Astronomy
Deep Survey which covered 2.5 deg$^2$ (as shown by the dotted lines)
from September 2001 to April 2002.    The $I_{AB} = 27$ limit in the
bottom panel corresponds to that of the ongoing COSMOS survey, which 
will survey an area of 2 deg$^2$ (again indicated by the dotted lines).
Note however, that the COSMOS survey itself is
primarily focused on large-scale structure issues and will not be
able to find PPSNe, as each pointing is visited only once.}
\label{fig:Iband}
\end{figure*}

\section{Pair-Production Supernovae in Cosmological Surveys}

From the models developed in \S 3, it is relatively straightforward to
relate the star formation history of VMS to the resulting number of
observable pair-production supernova.  In this section and below we
adopt cosmological parameters of $h=0.7$, $\Omega_m$ = 0.3,
$\Omega_\Lambda$ = 0.7, and $\Omega_b = 0.045$, where $h$ is the
Hubble constant in units of 100 km s$^{-1}$ and $\Omega_m$,
$\Omega_\Lambda$, and $\Omega_b$  are the total matter, vacuum, and
bayonic densities  in units of the critical density (\eg Spergel \etal
2003).

\begin{figure*}[t]
\vspace{145mm}
\includegraphics{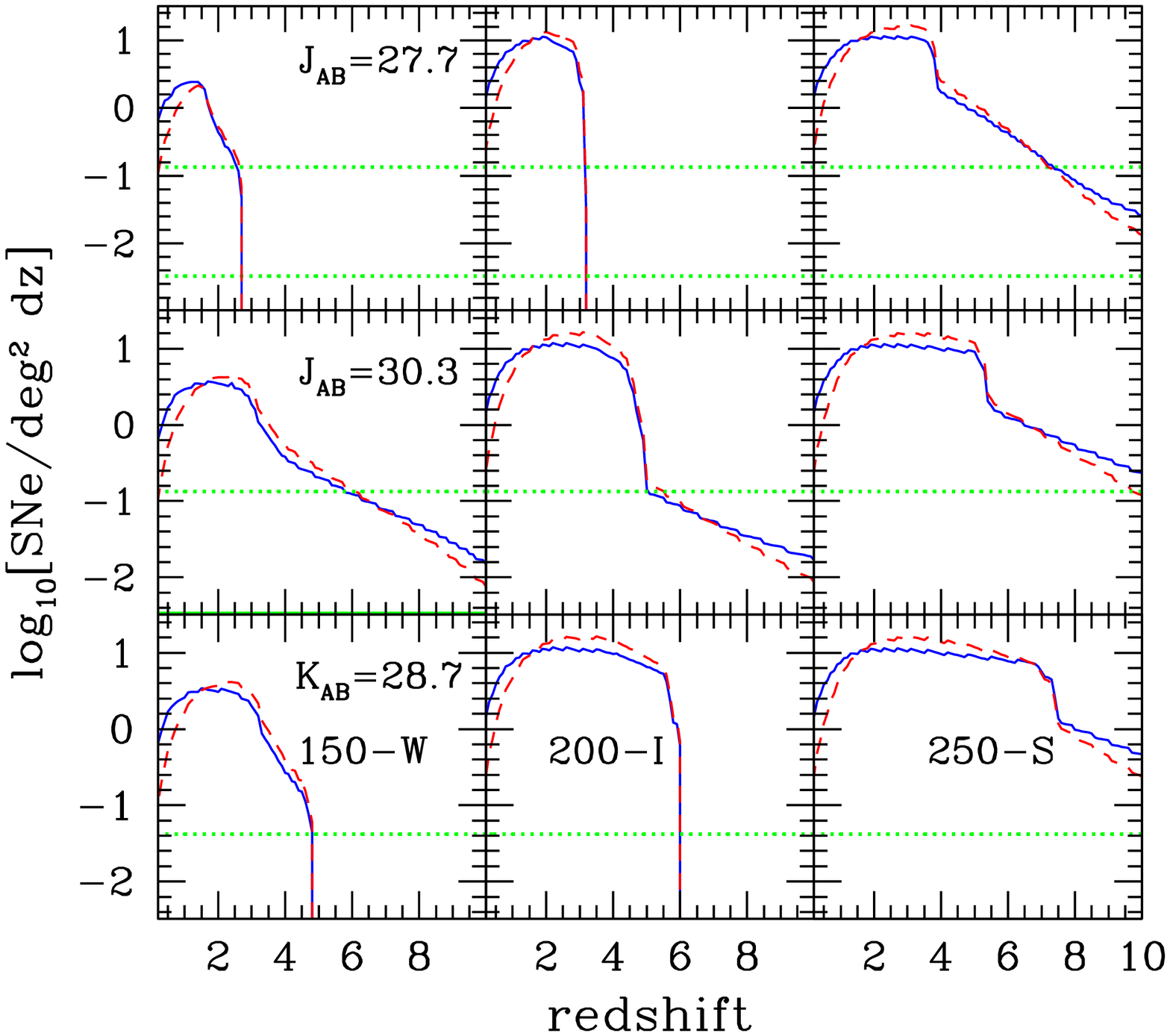}
\caption{\footnotesize 
Number of \sngg  per square degree per unit redshift above a
fixed J or K-band magnitude, with limits appropriate for three possible
realizations of the  {\em Joint Dark Energy Mission}.  Curves are as in
Figure \protect\ref{fig:Iband}.  Near infrared coverage allows for
detection of \sngg even after the lightcurves have  been significantly
redshifted.  {\em Top:} The $J_{\rm AB} = 27.7$ limit taken in these
panels corresponds to that of a single scan in the planned deep-field
{\em SNAP} survey, which would cover  7.5 deg$^2$ (indicated by the
upper dotted lines).   A similar magnitude limit with the same
coverage would be obtained with 2 months of grism data from the {\em
Destiny} mission.  The full planned 700 deg$^2$ wide-field {\em SNAP}
survey  (lower dotted lines) also has the same limiting magnitude.
{\em Center:} The $J_{\rm AB} = 30.3$ limit taken in these panels is that of
the deep-field {\em SNAP} survey, which would be able to  place
constraints on \sngg out to $z \gtrsim 5.$ {\em Bottom:} Curves
corresponding to the {\em JEDI} mission, whose redder photometry
coverage allows for detections down to $K_{\rm AB} = 28.7$ over 24
deg$^2$ with 2 months of data.  This would place limits on
\sngg out to $z \sim 6$ and beyond.}
\label{fig:Hband}
\end{figure*}

Here we focus on three PPSN light curves, which bracket the range of
possibilities: the faintest of all our models, 150-W,  in which both
significant dredge-up and  $^{56}$Ni production  are absent; an
intermediate model, 200-I, in which little dredge-up occurred, but 5.1
$\msun$ of $^{56}$Ni were formed; and the model with the brightest
lightcurves, 250-S, in which substantial dredge-up leads to an
enormous initial radius of over 20 A.U.,  and the production of 24.5
$\msun$ of $^{56}$Ni causes an extended late-time period of high
luminosity.  Accounting for redshifting, time dilation, and the
appropriate luminosity and bandwidth factors, the specific flux for
each of these models at a redshift $z$ observed at a wavelength
$\lambda$ and a time $t$ after  breakout (in the frame of the
observer) is 
\be 
F_\nu(\lambda,t,z) =  \frac{L_\nu \left[
\lambda (1+z)^{-1},t (1+z)^{-1} \right]} {4 \pi r(z)^2 (1+z)},
\label{eq:flux}
\ee 
where $r(z)$ is the comoving distance associated with the
supernova redshift.  Note that this expression does not address the
possibility of extinction by dust, which amounts to assuming that
pristine regions remain dust-free thought the lifetime of the very
massive PPSN progenitors stars.  For any given PPSN model, we can
then use eq.\ (\ref{eq:flux}) to calculate $t(\lambda, F_\nu^{\rm
min}, z)$  the total time the observed flux at the wavelength
$\lambda$  from a SN at the redshift $z$ is greater than the magnitude 
limit associated with the 
specific flux $F_\nu^{\rm min}.$ Finally, the total number of
pair-production SNe  shining at any given time with fluxes above
$F_\nu^{\rm min}$, per square degree per unit redshift is given by
the product of the volume element, the (time-dilated) PPSN rate
density, and the time a given PPSN is visible, that is
\be
\frac{dN_{\rm deg^2}}{dz} = 
[r(z) \, \sin(1 {\rm deg})]^2 \, \frac{dr}{dz} \frac{\rho^{\pm}(z)}{1+z}
t(\lambda, F_\nu^{\rm min}, z),    
\ee  
where the rate density $\rho^{\pm}(z)$ is the
number of \sngg  per unit time per comoving volume as a function of
redshift.

As $\rho^{\pm}(z)$ is completely unknown, we adopt here two simple
models.  In the first model, we  assume that metal-free star formation
occurs at a constant rate density, which we take to be $0.001 \, \msun
\, {\rm yr}^{-1} \, {\rm Mpc}^{-3},$  independent of redshift.   In
the second case, we assume that  at all redshifts metal-free stars
form at 1\% of the observed total star formation rate density, which
we model as $\log_{10}[\rho^{\rm obs}_\star(z)/ \msun \, {\rm yr}^{-1}
\, {\rm Mpc}^{-3}]  = -2.1 + 3.9 \, \log_{10}(1+z)  - 3.0 \,
[\log_{10}(1+z)]^2,$ a simple fit to the most recent measurements
(Giavalisco \etal 2004; Bouwens \etal 2004).  Finally, for both star
formation models we assume that  1 pair-production SN occurs per 1000
solar masses of metal  free stars. This is consistent with typical
estimates given in SSF03,  in  which $N^{\pm}$, defined as the number
of \sngg per solar mass of stars formed, was found to vary between
$\sim 10^{-4}$ and $\sim 5 \times 10^{-3}$ for a wide range of
possible metal-free IMFs.

Note that while our two simple PPSN rate densities are chosen such
that they can  be easily rescaled by the reader, they are
nevertheless consistent with the range of $z \gtrsim 1$ values
predicted in more sophisticated models, as discussed in \S 5.
Furthermore, the total amount of metals produced in our simple models
is consistent with the element abundances observed in extremely
metal-poor Galactic halo stars.  Assuming a typical value of 200 $\msun$
of metals ejected per PPSN and integrating down to $z=2.5$, one obtains
values  $\sim 6 \times 10^5 \msun$ Mpc$^{-3}$ for both of our PPSN
rate density models.  This corresponds to a mass fraction of baryons
that have been processed by VMS of $\sim 6 \times 10^{-5}$, which is
consistent with the $3-7 \times 10^{-5}$ limits inferred by Oh \etal
(2001) to explain the relative abundances in extremely metal-poor
Galactic stars.

The resulting observed \sngg counts for these models 
are given in Figure \ref{fig:Iband} for two
limiting magnitudes.  In the upper panels, we take a $I_{\rm AB} = 26$
magnitude limit, appropriate for the Institute for Astronomy (IfA)
Deep Survey (Barris \etal 2004), a  ground-based survey that covered a
total of 2.5 deg$^2$ from September  2001 to April 2002.  As we are
interested in rare objects, this type of survey is  more constraining
that a more detailed, smaller-area searches such as the  {\em Hubble}
Higher $z$ Supernova Search (Riess \etal 2004; Strolger \etal 2004).

From this figure we see that existing data sets, if properly analyzed,
are easily able to place useful constraints on VMS formation at low
redshifts.  Given a typical PPSN model like 200-I for example, the
already realized IfA survey can be used to  place a constraint   of
$\lesssim 1 \%$ of the total star formation  rate density out to a redshift
$\sim 1$.  Similarly, extreme models such as 250-S can be probed out
to redshifts $\sim 2$, all within the context of a recent SN search
driven by completely  different science goals.  Note however that
these limits are strongly dependent on significant mixing in the SN
progenitor or the production of $^{56}$Ni, and thus models such  as
150-W remain largely unconstrained by the IfA survey.

In the bottom panels of Figure \ref{fig:Iband} we consider a limiting
magnitude of  $I_{\rm AB} = 27$, appropriate for the COSMOS
survey\footnote{see http://www.astro.caltech.edu/$\sim$cosmos/},  an
ongoing project that will cover 2 deg$^2$ using the {\em Advanced
Camera for Surveys} on HST.  Raising the limiting magnitude from
$I_{\rm AB} = 26$ to $I_{\rm AB} = 27$ has the primary effect of
extending the sensitivity out to slightly higher redshifts.  This
pushes the probed range from $z \lesssim 1$ to $z \lesssim 1.5$ in the
200-I case and from  $z \lesssim 2$ to $z \lesssim 3$ in the 250-S
case.  Again this is all in the context of an ongoing survey.    Even
with this fainter limiting magnitude, however, low-luminosity \sngg
like 150-W  are extremely difficult to find, and remain largely
unconstrained.

This shortcoming is easily overcome by moving to NIR wavelengths.  In
Figure \ref{fig:Hband}, we calculate the PPSN constraints that would
be obtained from  three possible realizations of the planned
space-based  {\it Joint Dark Energy Mission (JDEM)}.   In the upper
and central panels of this plot we adopt two limiting $J$-band
magnitudes,  appropriate for the two surveys that would be carried out
with the  {\it Supernova Acceleration Cosmology Probe} ({\it
SNAP})\footnote{see http://snap.lbl.gov/}   realization  of {\em
JDEM}.   In the upper panel we adopt a limit of  $J_{\rm AB} = 27.7$,
corresponding to the depth of the planned (700 deg$^2$) wide-field
{\em SNAP} survey.   With 2 months of data, the alternative  {\em Dark
Energy Space Telescope} ({\em Destiny})\footnote{see
http://destiny.asu.edu/} realization of {\em JDEM} would cover 7.5
deg$^2$ of sky with {\em spectroscopic} observations down to a similar
magnitude limit.  

In these NIR surveys, ${dN_{\rm deg^2}}/{dz}$ is dramatically  
increased  with respect to ground-based
searches.   This is due to the fact that for the majority of their
lifetimes, the effective temperatures of \sngg are just above the
$\sim 10^{3.8} K$ recombination temperature of hydrogen, which
corresponds to a peak black-body wavelength  $\sim 8000$ \AA.   This
means that for all but the lowest redshifts, the majority of the
emitted light is shifted substantially redward of the I-band, which is
centered at $9000$ \AA.    Thus moving to the J-band, which is
centered at $12500$ \AA, represents an exponential increase in the
observed flux and allows for detections at significantly higher
redshifts.

In the 200-I case, this results in number densities that are almost
an order of magnitude higher than in the I-band.  Combined with the
larger area surveyed, this means that the wide-area {\em SNAP} survey
could place constraints on this  model that are roughly 300 times more
stringent than those from the IfA survey, and these constraints would
extend to $z \sim 4.$ Similarly tight limits can be placed on the
250-S model, but this time extending out to a redshift of 6.  In fact,
even very faint \sngg like 150-W would be sensitively probed  out to
$z \sim 2.5.$   In the central panels we adopt a limit of $J_{\rm
AB} = 30$, corresponding to the depth of the planned (7.5 deg$^2$)
{\em SNAP} deep-field survey.  In this case, both the 200-I and 250-S
models would be well studied out to  $z \gtrsim 5,$ and the 150-W model
would be easily detectable out to $z \sim 4$ despite its overall low
luminosity and relatively short lifetime.

Finally, in the lower panel, we consider an even redder 24 deg$^2$
survey with a limiting magnitude of $K_{\rm AB} = 28.7,$ as
appropriate for two months of observations from  the {\em Joint
Efficient Dark-energy  Investigation}   ({\em JEDI})\footnote{see
http://jedi.nhn.ou.edu/}  realization of {\em JDEM}.  As  the $K$ band
is centered at $22000$ \AA,  such observations naturally push to even
higher redshifts.   Thus while  reaching a limiting AB magnitude only
slightly higher than the  $700$ deg$^2$ {\em SNAP} survey, such a 24
deg$^2$ is able to place the most stringent of any of the surveys
considered: constraining the 150-W model out to $z \sim 4$ and the
more luminous models to $z \sim 6$ and beyond.

\section{The Environments of Pair-Production Supernovae}

As cosmological enrichment is essentially a local process, certain
environments are naturally more favorable for metal-free star
formation.  In SSF03, we showed that the transition from metal-free to
Population II stars was heavily dependent on the efficiency with which
metals where mixed into the intergalactic medium.   This efficiency
depended in turn on the energy input into galactic outflows powered by
\snggo, which was parameterized by the ``energy input per unit
primordial gas mass''  ${\cal E}_{\rm g}^{III}$, defined as  the
product of the fraction of gas in each primordial object that is
converted into stars ($f_\star^{III}$), the number of \sngg per unit
mass of metal-free stars formed (${\cal N}^{\pm}$), the average
kinetic energy per pair-production supernova (${\cal E}_{\rm kin}$),
and the fraction of the total kinetic energy channeled into the
resulting galaxy outflow ($f_{\rm wind})$.

Incorporating such outflows into a detailed analytical model of
structure formation leads to the approximate relation that, by mass,
the fraction of the total star formation in metal-free stars at $z=4$
is  \be F_\star^{III}(z=4) \sim 10^{-5}  ({\cal E}_{\rm
g}^{III})^{-1},  \ee  where, as above,  ${\cal E}^{\pm}$  is in units
of $10^{51}$ ergs  per $\msun$ of gas (see Figure 3 of SSF03 for
details).   Extrapolating the results in SSF03 to $z=0$ gives  \be
F_\star^{III}(z=0) \sim 10^{-5.5}  ({\cal E}^{III}_{\rm g})^{-1}.
\ee  These fractions can be related to the underlying population of
stars by adopting fiducial values of $f_\star^{III} = 0.1$ for the
star formation efficiency, which is consistent with the observed  star
formation rate density at intermediate and high redshifts (Scannapieco,
Ferrara, \& Madau 2002); $f_w = 0.3$ for the wind efficiency,  which
is consistent with the dwarf galaxy outflow simulations  of Mori,
Ferrara, \& Madau (2002); and $N^{\pm} = 0.001$,  for the number of
\sngg per unit solar mass of primordial stars formed, which is the
value assumed in \S 4.    This gives $F_\star^{III}$ values of  $0.3
({\cal E}_{\rm kin})^{-1}$ at $z=4$ and  $0.1 ({\cal E}_{\rm
kin})^{-1}$ at $z=0,$ respectively.   Or, in other words, for typical
energies of $30 \times 10^{51}$ ergs per \snggo, $\sim 1\%$ of the
star formation at $z=4$ and  $\sim 0.3 \%$ of the star formation at
$z=0$ by mass could be in metal-free.

In Figure 6 we show estimates of the number of SNe per
deg$^2$ per dz {\em per year} over the wide range of models considered
in SSF03, extrapolating to $z=0.$  In all cases we assume that 1
PPSN forms per 1000 solar masses of metal-free stars, and for
comparison we show the simple low-redshift  estimates taken in the
previous section.  While the values in these models are uncertain,
they nevertheless

{\medskip
\epsscale{1.02}
\plotone{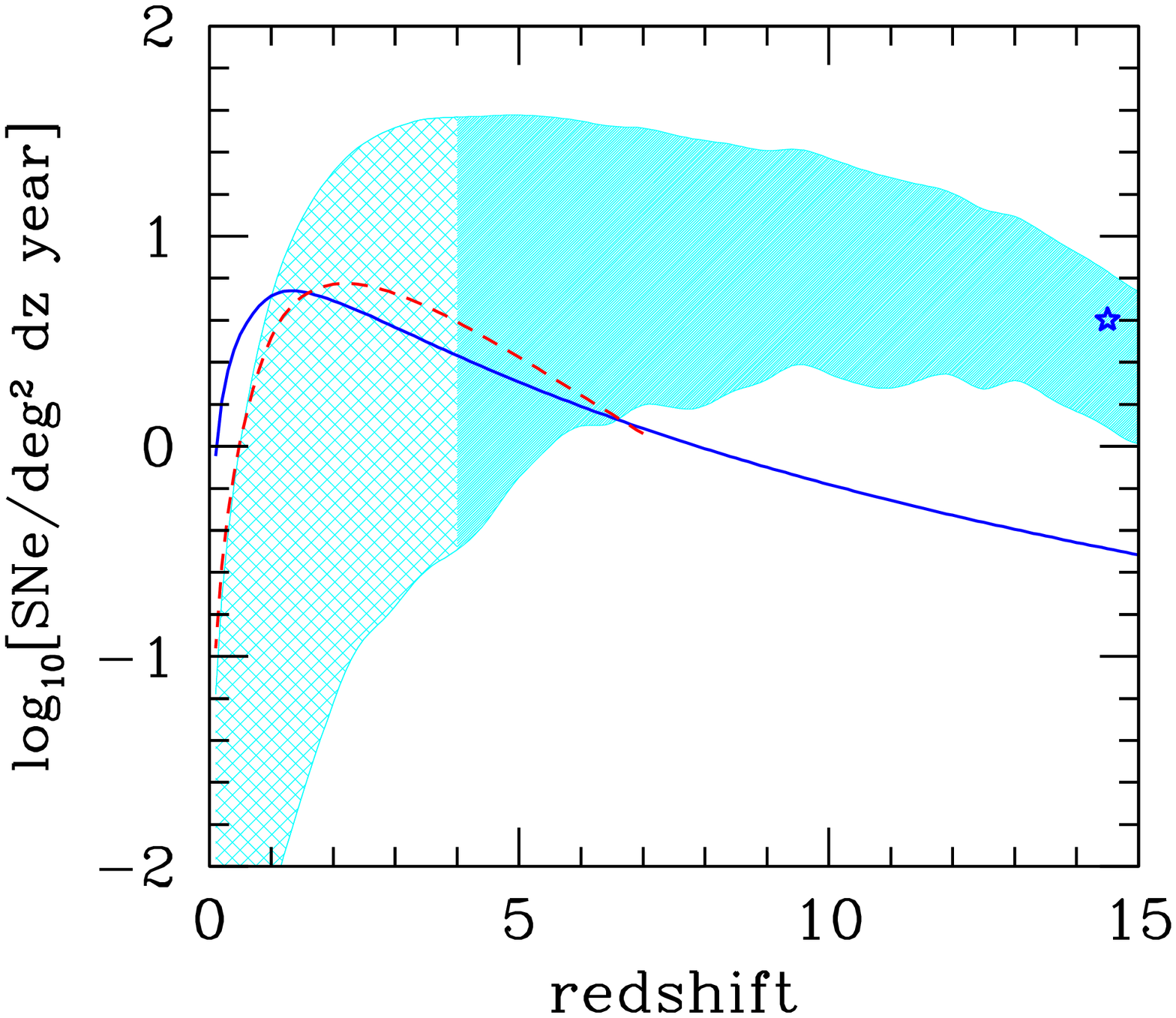}} 
{\footnotesize  {\sc Fig.}~6.---
Number of \sngg per square degree per unit redshift 
per year for a wide range of models.  As in Figs.\ 4 and 5,
the solid and dashed curves assume Pop III star formation
rate densities of 0.001 $\msun$ yr$^{-1}$ Mpc$^{-3}$ and 1 $\%$ of
the observed star formation rate density, respectively.   
The shaded region covers  the range of metal-free star formation 
rate density models 
considered in SSF03, with the weakest feedback
model (${\cal E}^{III}_{\rm g} = 10^{-4}$) defining the
upper end, and the strongest feedback model
(${\cal E}^{III}_{\rm g}= 10^{-2.5}$) defining the lower end.
An extrapolation of these star formation rate densities to
$z=0$ leads to the crosshatched region.
In all SSF03 models the highest rates occur at redshifts $\leq 10.$ 
Finally, the starred point is the $z=15$ 
estimate by Weinmann \& Lilly (2005). \medskip \medskip}

\noindent serve to illustrate two important points.   First,
even at the low redshifts probed by current surveys, the simple range
of VMS formation rates considered in \S4 lie well within  the
theoretically interesting range. Indeed for the weakest feedback cases
several $z \sim 1$ \sngg should have already been seen within the  IfA
Deep survey.  Secondly, due to the decrease in $dr$/$dz$ as a function
of redshift, as well as the time dilation effect, the peak of $dN_{\rm
deg}/{dz dt}$ is not in the $z \gtrsim 10$ range targeted by the {\em
James Webb Space Telescope}, but rather at more moderate redshifts.
This means that the constraints obtained from {\em
JDEM}-type surveys are likely to reach comparable limits to
higher-redshift efforts.  In fact, it is extremely unlikely that \sngg
will be found at higher redshifts without similar detections at
$z \lesssim 6.$


Note that for the full range of models in Figure 6,
metal-free star
formation naturally occurs in the smallest galaxies, just large enough
to overcome the thermal pressure of the ionized IGM, but small enough
not to be clustered near areas of previous star formation (SSF03).  In
our adopted cosmology, for an IGM temperature of $2 \times 10^4$, the
minimum virial mass is $8.4 \times 10^9 (1+z)^{-3/2} \msun$ with a
corresponding gas mass of $1.4 \times 10^9 (1+z)^{-3/2} \msun.$ This
means the total stellar mass of primordial objects is likely to be
around $M_\star \sim 10^8 \msun (1+z)^{-3/2}$, many orders of
magnitude below $L_\star$ galaxies.  Thus in general blank-field
surveys should be the best method for searching for  \snggo, as
catalogs of likely host galaxies would be extremely  difficult to
construct.

Nevertheless, as VMS shine so brightly, a direct search for primordial
host galaxies is not a hopeless endeavor.  In particularly the lack of
dust in these objects and the large number of ionizing photons from
massive metal-free  stars leads naturally to a greatly enhanced Lyman
alpha luminosity.  Following SSF03 this can be estimated as   \be
L_\alpha = c_L (1-f_{\rm esc}) Q(H) M_\star, \ee where $c_L \equiv
1.04 \times 10^{-11}$ ergs, $f_{\rm esc}$ is the escape  fraction of
ionizing photons from the galaxy, which is likely to be $\lesssim 0.2$
(see Ciardi, Bianchi, \& Ferrara 2002 and references  therein), and
the ionizing photon rate $Q(H)$ can be estimated as $\approx 10^{48}$
s$^{-1}$ $\msun^{-1}$ (Schaerer 2002).  This gives a value of
$L_\alpha  \sim 10^{45} (1+z)^{-3/2} {\rm ergs \, s}^{-1}$  which, if
observed in a typical $~1000$ \AA \, wide broad band corresponds to an
absolute AB mag $\sim -23 + 3.8 \log(1+z),$ much brighter than the
\sngg themselves.  However this flux would be spread out over many
pixels and be more difficult to observe against the sky than the
point-like \sngg emission.  For further details on the detectability of 
metal-free stars  though Lyman-alpha observations, the reader is 
referred to SSF03.

Finally we note that the presence of pair-production supernovae is
generic to stars that die within a given  range of helium core masses
(HW02), and does not directly depend on metallicity.    However there
are three important theoretical reasons that make metal-free stars the
favored progenitors of \snggo.  First, as $H_2$ becomes an inefficient
coolant below a typical density and temperature of $\sim 10^4$
cm$^{-3}$  and $\sim 100$ K, fragmentation to masses smaller than
$\sim 10^3 \msun$  is highly suppressed in gas of primordial
composition (\eg Abel, Bryan, \& Norman 2002).  Secondly,  metal-rich
VMS are prone to opacity-driven radial  pulsations, which are
suppressed in stars with metallicities below  $\sim 10^{-3} Z_\odot$
(Baraffe, Heger, \& Woosley 2001).  Lastly, the lack of metals greatly
reduces the line-driven wind mass loss, which is predicted to decline
with metallicity as $Z^{1/2}$ or faster (Kudritzki 2000; Vink \etal
2001; Kudritzki 2002).

From an observational point of view, Figer (2005) carried out a
detailed study of the IMF in the Arches cluster, which is large
($M_\star > 10^4 \msun$), young ($\tau$ = 2.0 - 2.5 Myrs), and at a
well-determined distance, making it ideal for such studies.  No stars
more massive than  130 $\msun$ were found in this $Z \sim Z_\odot$
cluster, although more  than 18 were expected.   A similar $\sim 150
\msun$ limit was found in the lower metallicity cluster R136 in the
Large Magellanic Cloud (Weidener \& Kroupa 2003), and while mass
estimates of a few stars exceed this limit, these values are quite
uncertain.  The Pistol star, for example, may in fact be a tight
binary or have recently experienced a merger with another star (Figer
\& Kim 2002).

Intriguingly, the Arches  cluster not only exhibits a cutoff  at the
highest masses but a flattened IMF above 50 $\msun$ ($\Gamma = -0.9$
rather than the $-1.35$ Salpeter slope).   Could some of these stars
have been quickly  whittled down from larger-mass progenitors?  The
answer is  unclear. What is clear, however, is that there is little
observational  or theoretical evidence that the progenitors of \sngg
might be found  in enriched environments.

\vspace{.1in}

\section{Summary}

Astronomers naturally associate metal-free star formation with
extremely high redshifts.  While the early universe contained no
elements  heavier than lithium, today stellar nucleosynthetic products
are found in all measured Galactic halo stars,
all nearby galaxies, and even in the low-density IGM.  Yet it would be
a mistake to conclude that such observations  exclude metal-free star
formation at moderate redshifts.  Metal-enrichment is an intrinsically
local process that proceeds over an extended redshift range, and at
each redshift, the pockets of metal-free star formation are naturally
confined to the lowest-mass galaxies, which are small enough not to be
clustered near areas  of previous star formation.   As such faint
galaxies are difficult to detect and even more difficult to confirm as
metal-free, the hosts of \sngg could easily be lurking at the limits
of present-day galaxy surveys.

Using the implicit hydrodynamical code KEPLER we have constructed a
suite of lightcurves that address the theoretical uncertainties
involved in modeling  \snggo.  Here the most important factors are the
mass of the progenitor star and the efficiency of dredge-up of carbon
from the core into the envelope.  In general, increasing the mass
leads to greater $^{56}$Ni production,  which boots the late time SN
luminosity.   Mixing, on the other hand, has two major effects: it
increases the opacity in the envelope, leading to a red giant phase
that increases the early-time SN luminosity; and it decreases the
mass  of the He core, consequently leading to a somewhat smaller mass
of $^{56}$Ni being synthesized.  Despite these uncertainties, \sngg in
general can be characterized by three key features: (1) peak
magnitudes that are brighter than Type II SNe and comparable or
slightly brighter than typical SNe Type Ia;  (2) very long decay times
$\sim 1$ year, which result from the  large initial radii and large
masses of material involved in the explosion;  and (3)  the presence
of hydrogen lines, which are caused by the  outer envelope.  Note,
however, only this last feature is  present in {\em all} cases, and in
fact,  the lowest mass PPSN models we constructed
have lightcurves that are remarkably similar to those of SN Type II.

Accounting for redshifting, time dilation, and appropriate  luminosity
factors we used these lightcurves to relate the overall very massive
star formation rate density to the number of \sngg detectable in
current and planned supernova searches.  Here the long lifetimes help
to keep a substantial number of \sngg  visible at any given time,
meaning that ongoing SN searches  should be
able to limit the contribution of VMS to  $\lsim 1\,$\% of  the total
star formation rate density out to a redshift of 2, unless both mixing and
$^{56}$Ni production are absent for all \snggo. Such
constraints already place meaningful  limits on the  cosmological
propagation of metals.

The impact of future NIR searches is even more promising, as the
majority of the PPSN light is emitted at restframe wavelengths
longward of $\sim 8000$ \AA.  Thus planned NIR satellite missions such
as {\em JDEM} would be over two orders of  magnitudes more sensitive
to \sngg than present optical surveys, and able to probe redshifts 
beyond $z \approx 6$.   In this case, even the dimmest \sngg would be
detectable out to $z \approx 4$.

Although the peak of the metal-free star formation density almost 
certainly occurred at extremely early times, there is much to be learned from
PPSN searches at more moderate redshifts. In fact, due to 
volume and time-dilation effects, the peak in the number of
\sngg per deg$^2$ per d$z$ per year is likely to lie well
below $z=10.$   Furthermore, the data sets
necessary for such analyses are already being planned for and
collected.  While the properties \sngg are diverse, a singular
conclusion can be drawn from our modeling.   Searches for
pair-production SNe at $z \lesssim 6$ will  dramatically increase our
understanding of the history of cosmic enrichment, the nature of
metal-free stars, and  the evolution of gaseous matter in the universe.

\acknowledgments

We thank Brian O'Shea \& Zoltan Haiman
for helpful discussions about primordial star
formation, Avishay Gal-Yam, Peter Garnavich, Weidong Li, Dovi
Poznanski,  Tony Spadafora, and Yun Wang, for information on current
and upcoming supernova surveys, and Nick Scoville for information
on the COSMOS survey.  We are also grateful to the
anonymous referee for helpful suggestions that greatly improved
the manuscript.  This work was supported by the National
Science Foundation under grants PHY99-07949, AST02-06111, and
AST02-05738, by NASA grants NAG5-11513, NAG5-12036 and NNG04GK85G, and
by the DOE Program for Scientific Discovery  through Advanced
Computing (SciDAC; DE-FC02-01ER41176).  PM acknowledges support
from the Alexander von Humbold Foundation. 
AH was also funded by DOE contract
W-7405-ENG-36 to the Los Alamos National Laboratory, by NASA grants
SWIF03-0047-0037 and NAG5-13700, and under NASA/STSci
HST-GO-09437.08-A.

\fontsize{10}{10pt}\selectfont

\end{document}